\documentclass[11pt]{article}

\usepackage{scicite}
\usepackage{graphicx}
\usepackage{subfigure}

\usepackage{times}

\newenvironment{sciabstract}{%
\begin{quote} \bf}
{\end{quote}}

\newcounter{lastnote}

\title{Charge Trap Memory Based on Few-Layered Black Phosphorus}

\author
{Qi Feng$^{\dagger}$, Faguang Yan$^{\dagger}$, Wengang Luo,  and Kaiyou Wang$^{\ast}$ \\
\\
\normalsize{SKLSM, Institute of Semiconductors, }\\
\normalsize{Chinese Academy of Science, Beijing 100083, China}\\
\normalsize{$^\ast$To whom correspondence should be addressed; E-mail:  kywang@semi.ac.cn}\\
\normalsize{$^\dagger$ These authors contributed equally to the work}
}

\date{}

\begin{document}

\baselineskip24pt

\maketitle

\begin{sciabstract}
 Atomically thin layered two-dimensional materials, including transition-metal dichacolgenide (TMDC) and black phosphorus (BP),\cite{1} have been receiving much attention, because of their promising physical properties and potential applications in flexible and transparent electronic devices . Here, for the first time we show non-volatile charge-trap memory devices, based on field-effect transistors with large hysteresis, consisting of a few-layer black phosphorus channel and a three dimensional (3D) Al$_2$O$_3$ /HfO$_2$ /Al$_2$O$_3$ charge-trap gate stack. An unprecedented memory window exceeding 12 V is observed, due to the extraordinary trapping ability of HfO$_2$. The device shows a high endurance and a stable retention of $\sim$25$\%$ charge loss after 10 years, even drastically lower than reported MoS$_2$ flash memory. The high program/erase current ratio, large  memory window, stable retention and high on/off current ratio, provide a promising route towards the flexible and transparent memory devices utilising atomically thin two-dimensional materials.  The combination of 2D materials with traditional high-$k$ charge-trap gate stacks opens up an exciting field of nonvolatile memory devices. 

\end{sciabstract}

\section*{Introduction}

Nonvolatile memory cells, which are essential components for digital, portable and self-standing electronics, attract more and more attention, since the miniaturization, low power consumption and reliable data storage are highly desirable to solve the problem of large data capacity, integration density. Memories based on ultrathin layered two-dimensional (2D) materials like graphene and transition-metal dichalcogenides (TMDCs), such as MoS$_2$, even graphene-oxide which can act as field effect transistor (FET) channel, charge-trapping layer or electrode, have been demonstrated and considered to be promising candidates, due to their fantastic electronic properties and potential applications \cite{1,2,3,4,5,6,7,8}. 
It was shown that the large hysteresis in the gate characterization curves of FETs can be applied for memory devices operation. Despite the extremely high mobility in graphene, the absence of a band gap hinders the achievement of a high ON/OFF ratio \cite{1,9}. This has led to intensive research in other 2D materials that have an intrinsic band gap. Among them, TMDCs, such as MoS$_2$ and WS$_2$, which have finite band gaps, not only have enabled the fabrication of high-performance field-effect transistor (FET) devices but have also paved the way for the realization of novel optoelectronic and valleytronic devices \cite{10,11,12,13,14,15}. 
Recently, black phosphorus monolayer, as well as its multilayers have been extensively studied \cite{16,17,18,19,20,21,22,23,24}. Comparing with  TMDCs which have relatively low mobility, black phosphorus (BP) becomes an outstanding new element 2D layered material, with high mobility value of 1000 cm$^2$V$^{-1}$s$^{-1}$ and 3900 cm$^2$V$^{-1}$s$^{-1}$ at room temperature and low temperature, respectively \cite{25,56}. BP is a van der waals type semiconducting layered material with a direct band gap of 0.3eV (bulk) to 2eV (monolayer), depending on the number of layers \cite{26}. The superior ambipolar property, higher mobility than TMDCs, comparable drain current modulation up to ~10$^5$ with TMDCs, promise black phosphorus as a good candidate for memory devices \cite{25,27,28,29,30}. Moreover, nonvolatile memory based on black phosphorus also hold great promise for future flexible and transparent devices because of the mechanical flexible and tunable electronic properties of black phosphorus \cite{31,32,33}.  

The selection of charge trapping layer and dielectrics is also significant because the  applications of nonvolatile memory are still constrained by low charge retention and high operating voltage. In recent years, different structures of the memory devices have been investigated, by using graphene, MoS$_2$ or high-$k$ oxide dielectric such as HfO$_2$, HfAlO, HfON and TiO$_2$ as the trapping layer \cite{2,3,5,34,35,36,37}. The related works indicate the high efficiency and potential application of HfO$_2$ used as charge trapping layer\cite{38,39,40,41}. Meanwhile, a high quality dense tunneling dielectric and high-$k$ blocking layer can provide low charge leakage and reduce the operating voltage as well as power consumption, respectively. Therefore, thanks to the high-$κ$ HfO$_2$ as the trapping layer while the Al$_2$O$_3$ of different thickness as the tunneling and blocking layer, the structure of Al$_2$O$_3$/HfO$_2$/Al$_2$O$_3$ (AHA) gate stack can be utilised effectively in nonvolatile memory to obtain devices with excellent properties, such as good trap ability, low power consumption and outstanding thermal stability \cite{3,40,41}. However, to our knowledge, there is no black phosphorus involved nonvolatile memory device being reported yet. We expect to enhance the performance of memory devices by combining the few-layer black phosphorus with conventional Al$_2$O$_3$/HfO$_2$/Al$_2$O$_3$ gate stack. 

In this paper, we demonstrate memory devices fabricated from black phosphorus and high-$k$ HfO$_2$. As expected, the device shows a significant hysteresis and a substantial memory window thanks to the superior trap capacity of the  Al$_2$O$_3$/HfO$_2$/Al$_2$O$_3$ (AHA) gate stack. Meanwhile, a robust charge retention of 70\% retain after 10 years and stable endurance of more than 1200s and 120 cycles are obtained. The application of the conventional Al$_2$O$_3$/HfO$_2$/Al$_2$O$_3$ gate stack renders a possibility for a massive production of high-performance of black phosphorus-based 2D memory devices.

\section*{Results and discussion}
Few-layer black phosphorus is obtained through mechanical exfoliation from  black phosphorus bulk crystals onto prepatterned SiO$_2$/Si$^{++}$ substrates with the thickness of SiO$_2$ of 300 nm \cite{25}. We chose the particular flake with a thickness of 15 nm, as confirmed by atomic force microscopy, as extremely thin flakes ( $<$5 nm) have a low mobility of $\sim$ 10 cm$^{2}$V$^{-1}$s$^{-1}$ \cite{17,43}.  Source and drain contacts of 5 nm/85 nm Ti/Au were deposited $via$ thermal evaporation. Subsequently, the Al$_2$O$_3$/HfO$_2$/Al$_2$O$_3$ (AHA) gate stack was grown $via$ an atomic layer deposition (ALD) system with layer thickness of 5/8/35 nm, respectively. More details of device fabrication process can be found in the Methods.

\begin{figure}[h]
	\centering
	\includegraphics[width=0.4\textwidth]{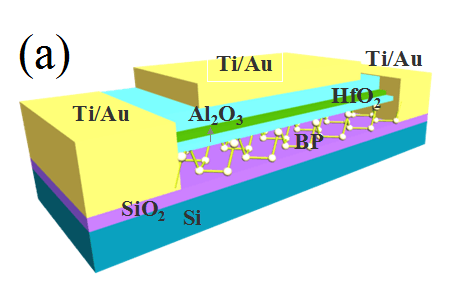}
	\includegraphics[width=0.4\textwidth]{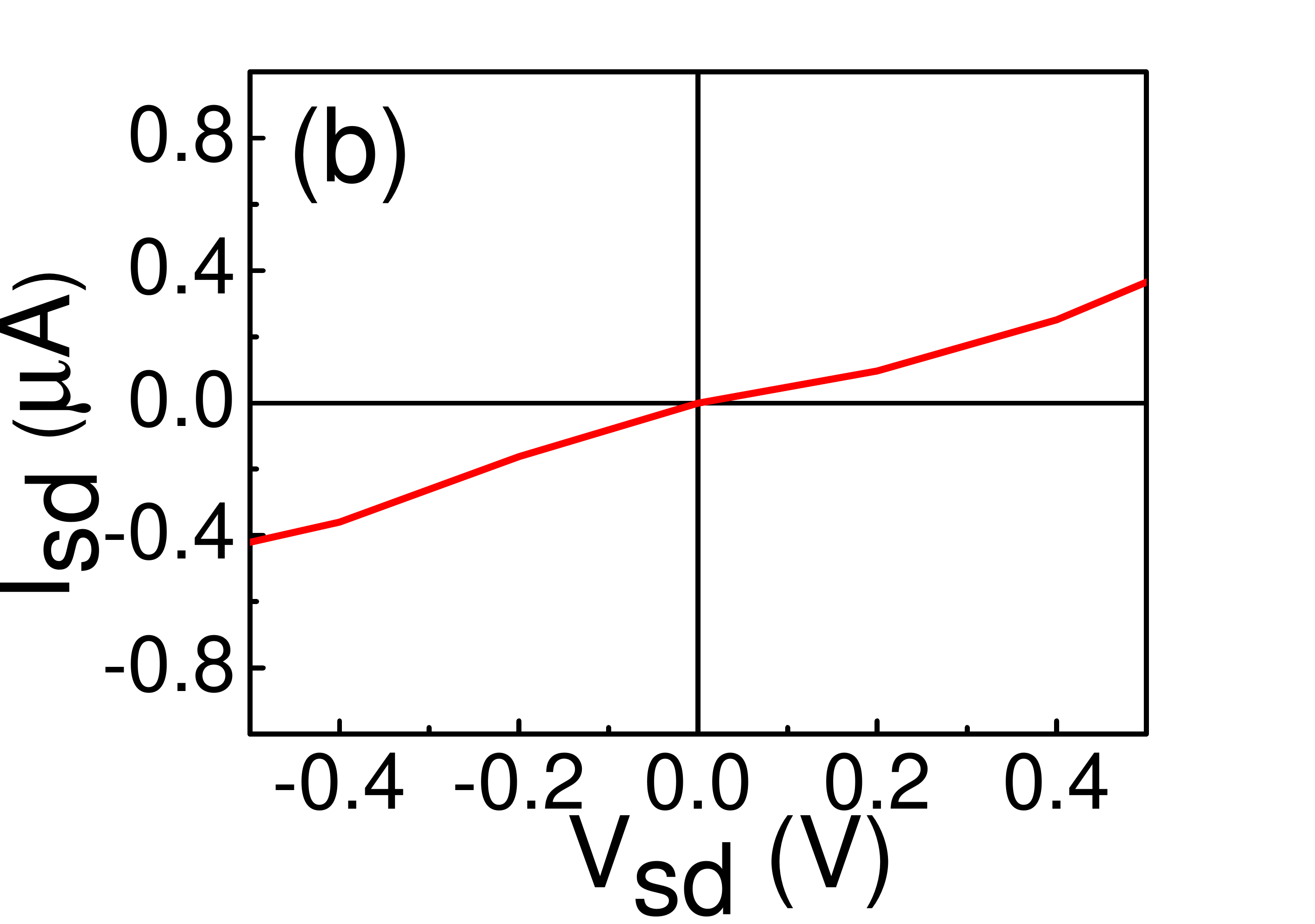}
	\includegraphics[width=0.4\textwidth]{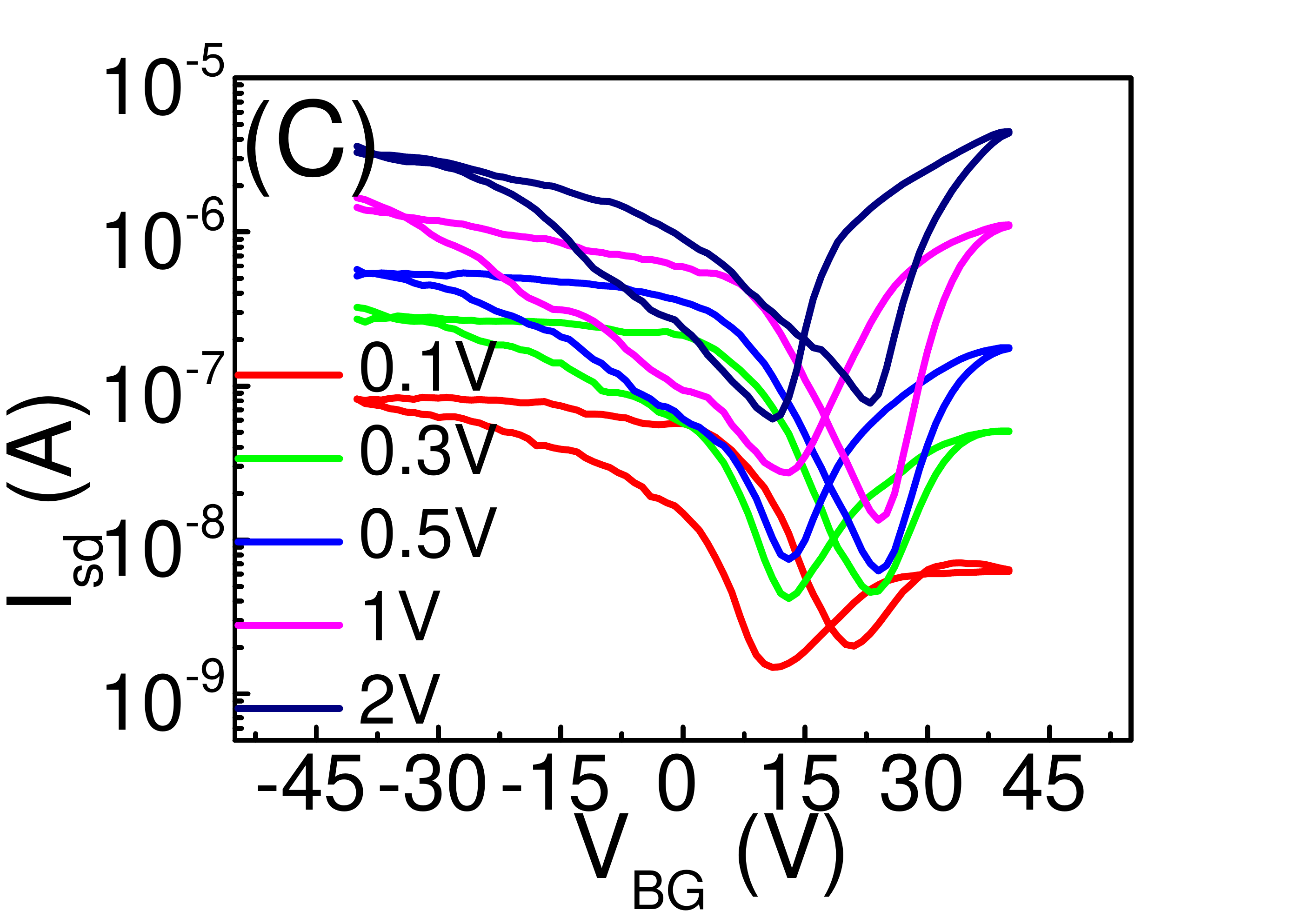}
	\includegraphics[width=0.4\textwidth]{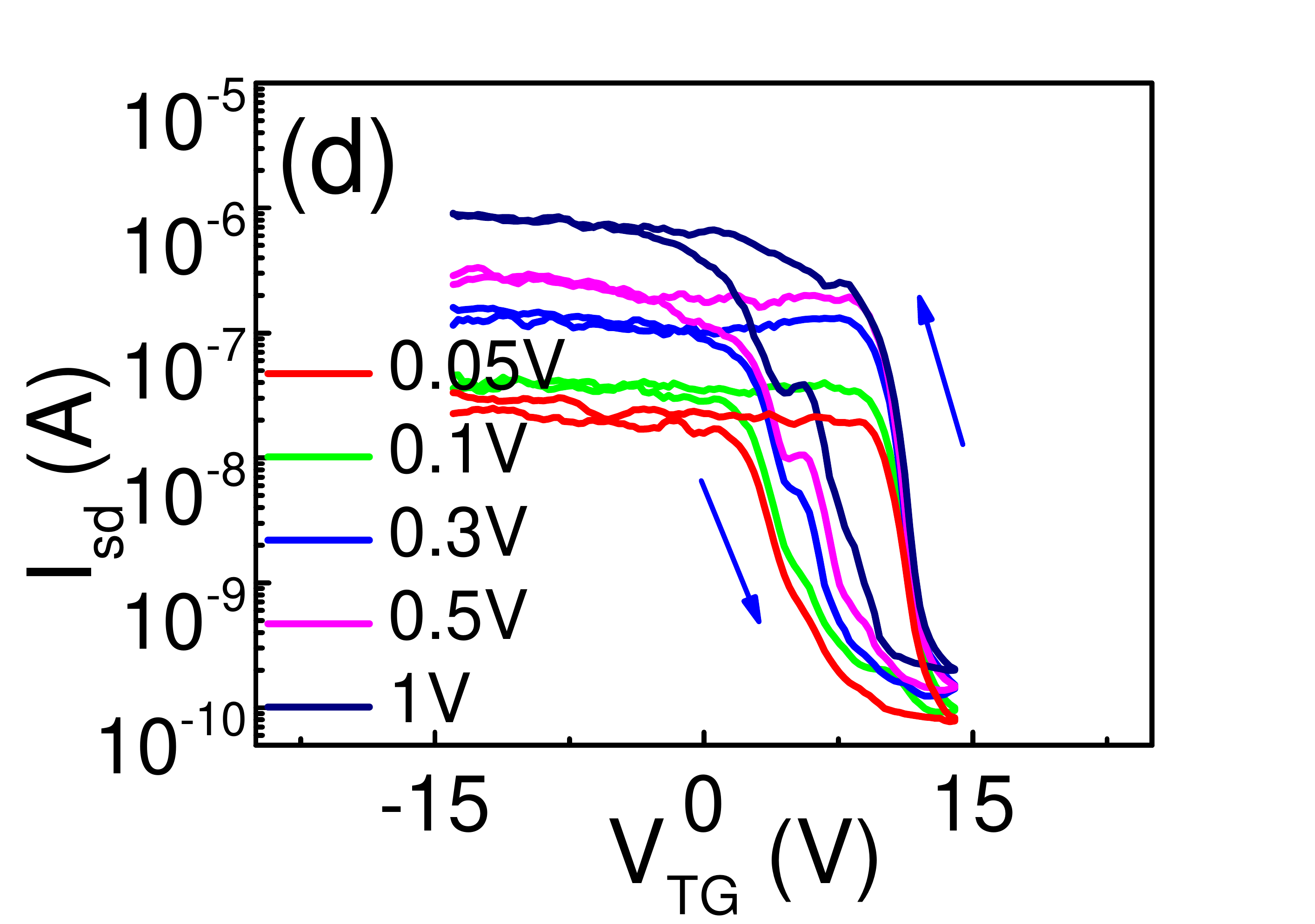}
	\caption{(Color Online) {\bf Black Phosphorus Device Characterisation} (a) Device structure. The few-layer BP and 8-nm HfO$_2$ serve as the channel and the trap layer, respectively. (b) The ouput characteristics (I$_{sd}$-V$_{sd}$) of the device in logrithm scale. Inset shows the IV curve in linear scale within smaller voltage scale. (c)  Transfer curves of the device with the back gate, sweeping between -40 V and 40 V,	 and (d) transfer  curves of the device with the  top gate, sweeping between -15 V and 15 V, with different source-drain voltage as denoted. The sweeping direction is denoted as blue arrows. }\label{fig:device}
\end{figure}

A typical device schematic is shown in Fig.\ref{fig:device}(a). The non-Ohmic contact was indicated in Fig.\ref{fig:device}(b) and sublinear dependence of I$_{sd}$ was clear in the small range as shown in the inset. An applied top-gate voltage (V$_{TG}$) modulates the amount of charge stored in the HfO$_2$ charge-trap layer, causing the variation of the conductivity of the BP channel. A back-gate voltage (V$_{BG}$) was applied to the degenerately doped silicon substrate to tune the memory characteristics by systematically shifting the Fermi level of black phosphorus. 
We noted that through the ALD deposition process the device performance can be significantly improved because of the thermal annealing under vacuum \cite{19,44}. Also, the encapsulation of black phosphorus in a high-$k$ dielectric environment will reduce the Coulomb scattering and modify the phonon dispersion in few-layer black phosphorus. Moreover, the oxide capping would protect the BP from ambient degradation \cite{45,46}.  
The black phosphorus in our device is measured to be about 15 layers by atomic force microscopy. The transfer curve  (I$_{DS}$-V$_{TG}$) of the device can be obtained by sweeping V$_{TG}$ while keeping the back gate grounded. When V$_{BG}$ is swept between -40 to +40 V, the appearance of the hysteresis between the forward and backward sweep curves indicates the interface effect from SiO$_2$ and the black phosphorus channel, as shown in Fig.\ref{fig:device}(c).

The transfer characteristics were  explored to probe the storage capability of the black phosphorus memory device.
As shown in Figure \ref{fig:device}(d), with V$_{BG}$ = 0 V and different V$_{sd}$ between 50mV and 1V, the transfer characteristic curves were obtained by sweeping V$_{TG}$ from -15V to +15V, then back to -15V, and a maximal on/off ratio about 10$^4$ was acquired.
A large memory window of about 12 V was observed, primarily originated from a large amount of electrons and holes stored in the charge-trap layer of HfO$_2$. The mobility degradation is likely due to the thickness of 15 nm of BP.  When different voltage sweep rates are applied, negligible appreciable differences are noted in transfer curves. It indicates that the hysteresis of the transfer curve is not caused by the captured molecules, such as a thin layer of water, at the interface of BP/Al$_2$O$_3$ \cite{47}. The observed large gate hysteresis can be utilized for a nonvolatile memory device operation employing the BP layer as the channel.

\begin{figure}
	\centering
	\includegraphics[height=5cm]{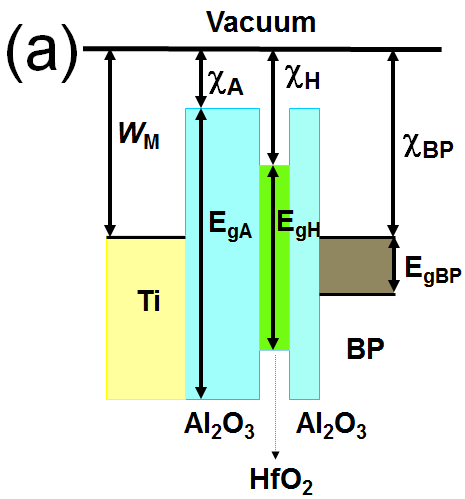}
	\includegraphics[height=5cm]{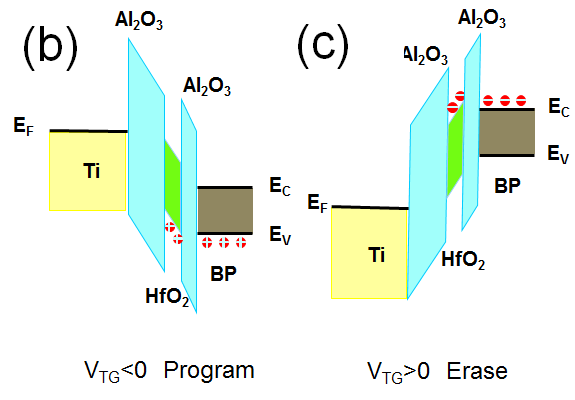}
	\caption{ (Color Online) {\bf Band diagram of BP memory device} (a) Band diagram of the BP-CTM memory cell with the AHA stack. (b)Band diagram of the program/erase state of the device under negative and positive V$_{TG}$. Negative V$_{TG}$ programs the device. Holes tunneling from the few-layer black phosphorus channel are accumulated in the HfO$_2$ charge-trap layer. Positive V$_{TG}$ erases the device. Electrons tunnel from the few-layer BP channel to the HfO$_2$ charge-trap layer.}\label{fig:band}
\end{figure}

The band diagram of the BP CTM with the AHA stack is shown in Figure \ref{fig:band}. The electron affinity of Al$_2$O$_3$ and HfO$_2$ are 1eV and 2.5 eV, and the band gap are 7.7 eV and 4.9 eV, respectively (as shown in Fig.\ref{fig:band}(a)). The electron affinity of BP is about 4.4 eV and the band gap of bulk and monolayer BP are 0.3 eV and about 1.0 eV, respectively. In this situation, electron and hole potential well will be formed in the AHA stack layers structure, which can trap electrons and holes.  In consideration of the 5 nm thick Al$_2$O$_3$, electrons can tunnel through the Al$_2$O$_3$ barrier by means of the mechanism of Fowler-Nordeim tunneling \cite{48}. In the device, the 5nm Al$_2$O$_3$-layer acts as tunnel layer, while 8nm HfO$_2$-layer acts as trap layer.

As presented in Fig.\ref{fig:band}(b,c), negative and positive gate voltage applied to the top gate correspond to the program and erase states in the device operation process, respectively. In the program process, the shape of the energy band changes as shown in Fig.\ref{fig:band} and the barrier of the tunneling oxide layer which is close to the BP film becomes thin enough for tunneling; hence, holes in BP layer can be pulled and transferred to HfO$_2$-layer through the Al$_2$O$_3$ tunnel layer as a result of the Fowler-Nordeim tunneling effect (see Fig.\ref{fig:band}). The resultant accumulation of holes in HfO$_2$ will shift the threshold voltage (V$_{th}$) to the negative direction, since this accumulation of holes in the HfO$_2$ layer would screen the top-gate electric field effect. When the gate voltage is swept toward a higher positive value, the holes in trap layer will be pushed back to the BP channel; meanwhile, the electrons in conduction band of BP will be pulled into the trap layer which also can screen the top-gate electric field and cause the threshold voltage shifted to the positive direction (see Fig.\ref{fig:band}). Then, an appreciable memory window can be obtained in the transfer characteristic curve as a result of the capability of electron and hole trapping from AHA stack. As the top-gate voltage changes from positive to negative, the memory cell works in the program state, while as the top-gate voltage changes from negative to positive, the memory cell works in the erase state.

\begin{figure}
	\centering
	\includegraphics[width=0.4\textwidth]{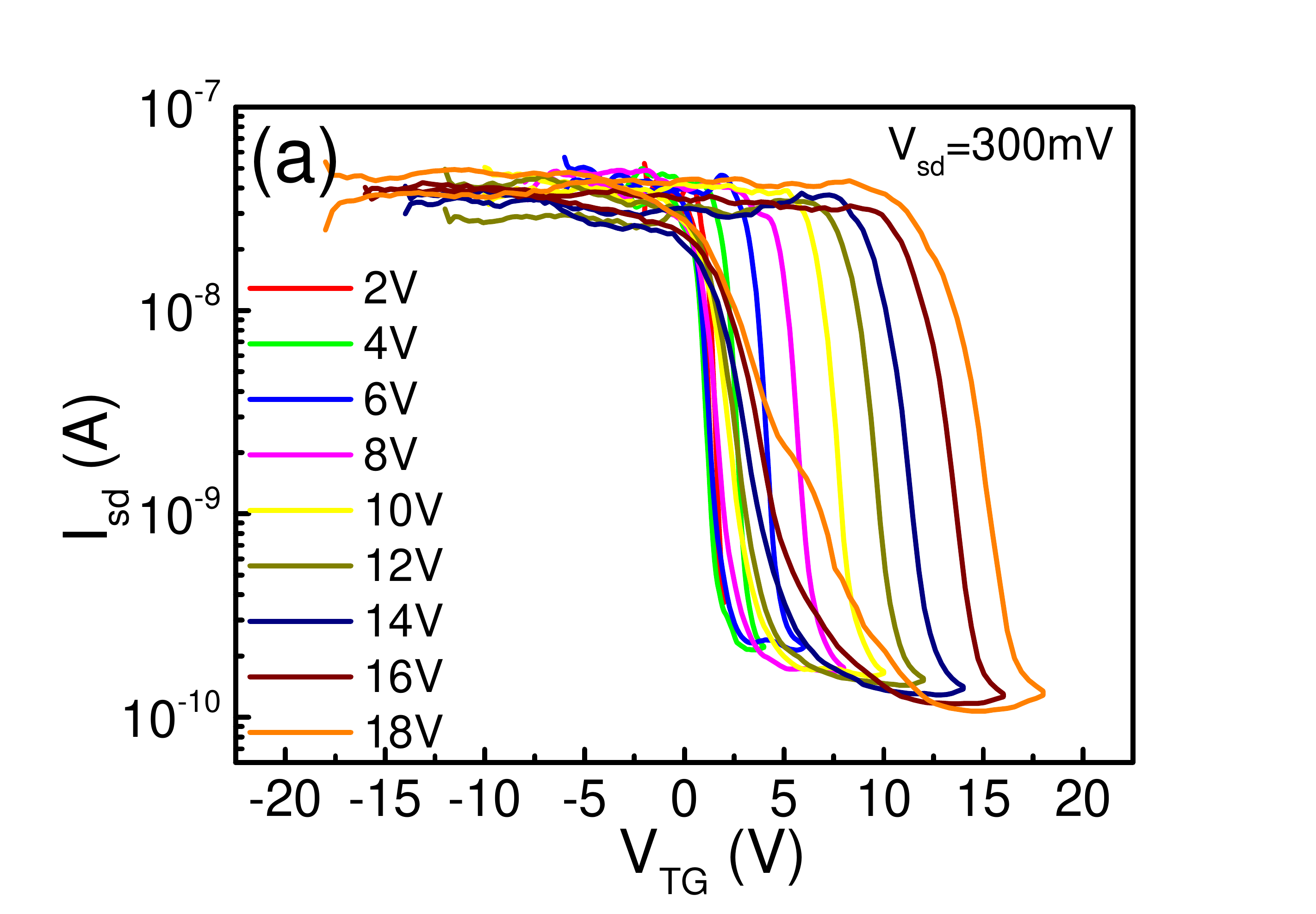}
	\includegraphics[width=0.4\textwidth]{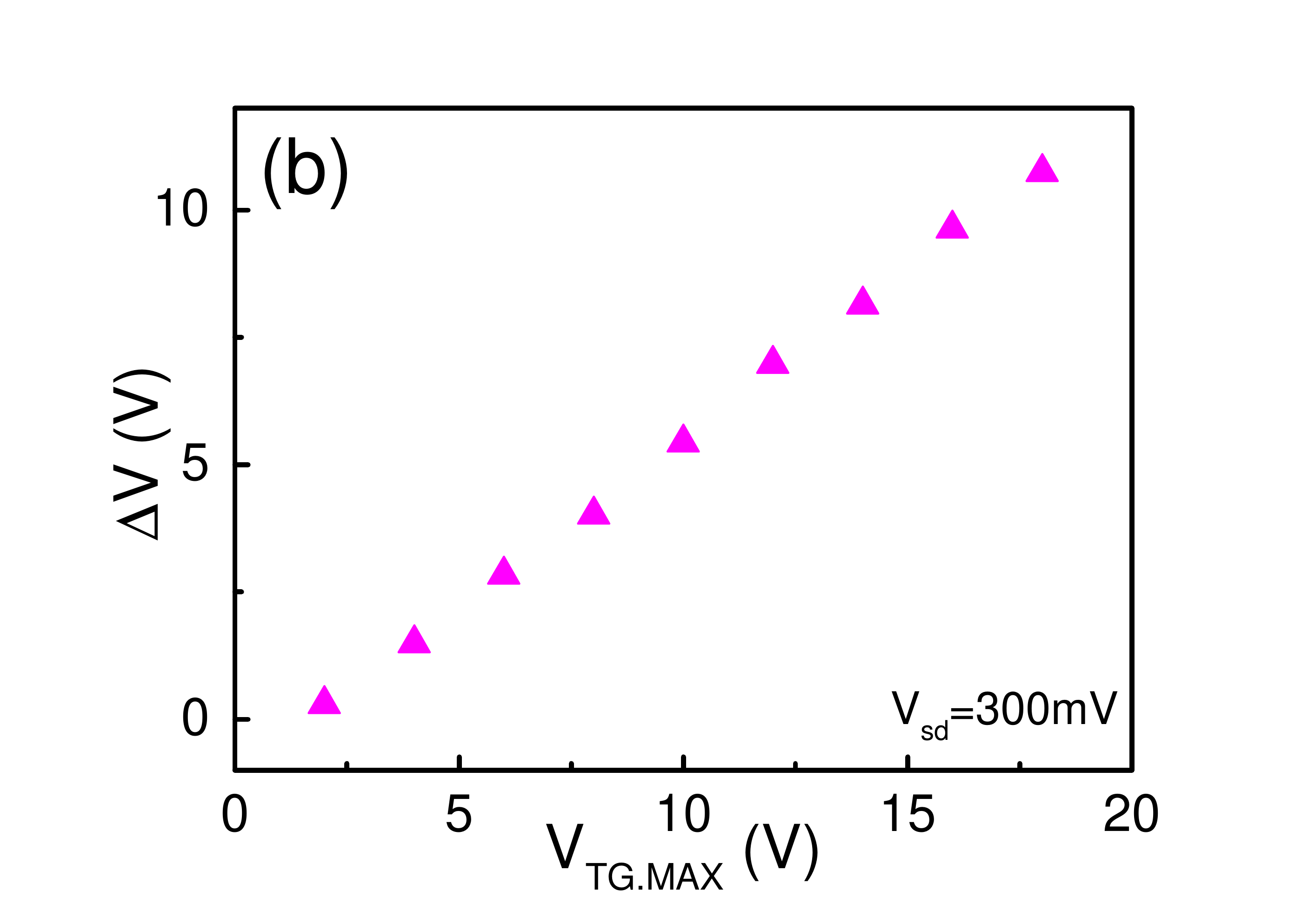}
	\caption{ (Color Online) {\bf Charge-Trap Memory Window} (a) I$_{sd}$-V$_{TG}$ characteristics under different V$_{TB.MAX}$ at V$_{sd}$=300 mV. (b) Extraction of memory window $vs$  V$_{TB.MAX}$.  The memory window increases from $\sim$ 1 to $\sim$ 12 V in our experimental settings.}\label{fig:window}
\end{figure}

The amount of charge stored in the charge-trap layer can be modulated by gradually changing the maximum (+V$_{TG.MAX}$) and minimum (-V$_{TG.MIN}$) voltage applied on the top gate. Figure \ref{fig:window}(a) shows the enlarged hysteresis window when $|V_{TG.MAX}|$ becomes larger. The shift of the threshold voltage toward negative and positive direction corresponds to the hole and the electron trapping, respectively. The increased threshold voltage shift ($\Delta V$) as a function of V$_{TG.MAX}$ is summarized in Figure \ref{fig:window}(b). The amount of charge stored in the charge-trap layer can be estimated from the expression:
\begin{equation}
	n = \frac{\Delta V C_{HF-AL}}{e} \label{eqn:n}
\end{equation}
where $e$ is the electron charge, $\Delta V$ is the threshold voltage shift toward the negative or the positive direction compare to the original transfer curve \cite{49}. According to Figure \ref{fig:window}(a), the hysteresis window of the sweep between -2 and +2 V is close to zero, thus it can be defined as the original transfer curve, corresponding to no tunneled electrons or holes existing in the charge-trap layer. $C_{HF-AL}=\epsilon_{0}\epsilon_{AL}/d_{AL}$ is the capacitance between the HfO$_2$ charge-trap layer and the top gate, where $\epsilon_{0}$ is the vacuum permittivity, $\epsilon_{AL}$ and $d_{AL}$ are the relative dielectric constant ($\sim$ 8) and thickness ($\sim$35 nm) of the Al$_2$O$_3$ blocking layer, respectively. 

Here, we define $\Delta$V =$\Delta$V$_h$ + $\Delta$V$_e$ , where $\Delta$V$_h$ and $\Delta$V$_e$ are the threshold voltage shift toward the negative and positive direction compare to the original transfer curve, corresponding to the density of stored holes and electrons, respectively. The calculated density of stored electrons and holes under |V$_{TG.MAX}$| of 18V is on the order of 10$^{14}$ and 10$^{13}$ cm$^{-2}$, respectively, which is in agreement with previously reported memory devices using graphene or graphene oxide as charge-trap layers \cite{5,7,8}. The lower tunneling barrier height of electrons than holes is accounted for the high trap density of electrons.

\begin{figure}
	\centering
	\includegraphics[width=0.4\textwidth]{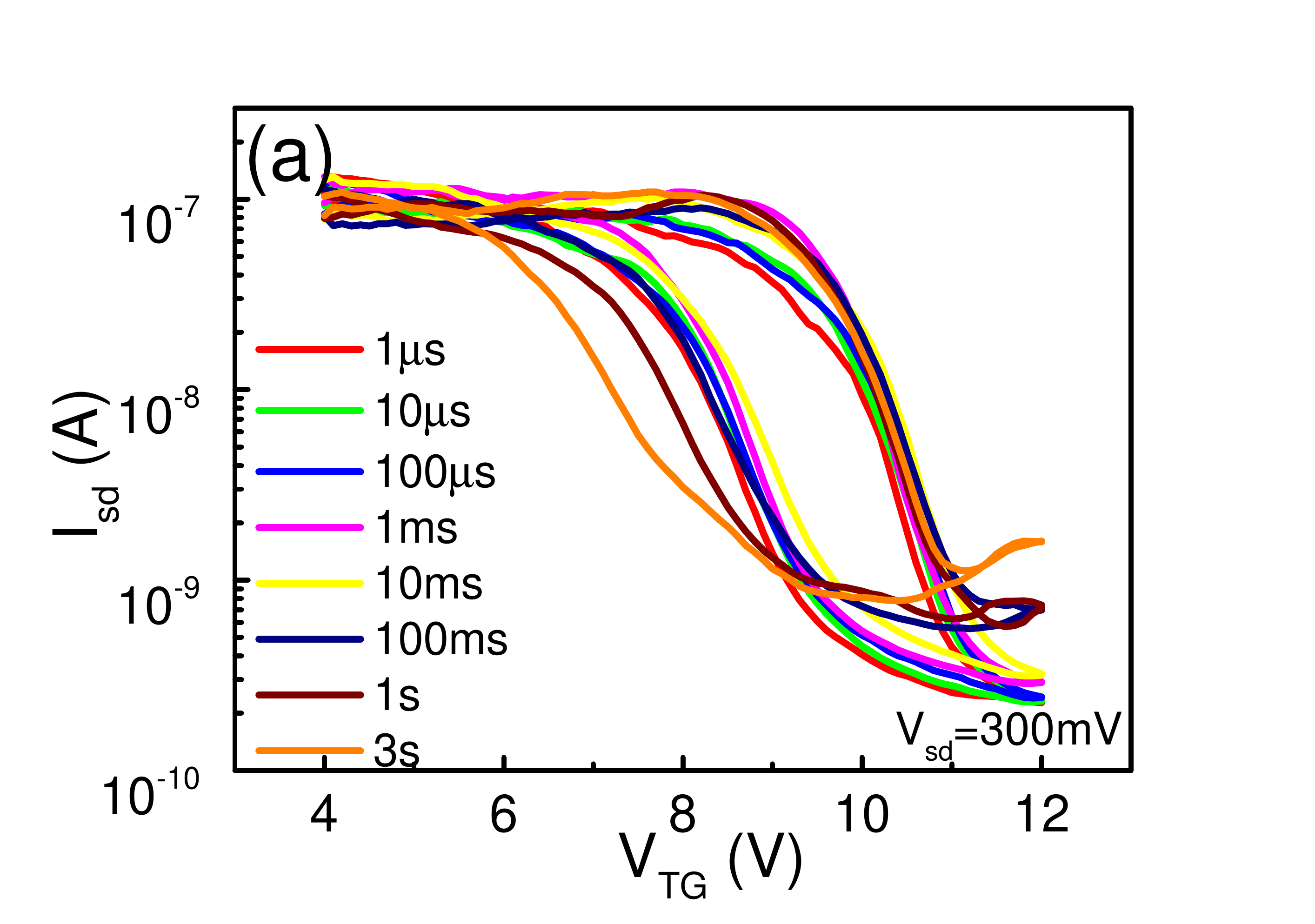}
	\includegraphics[width=0.4\textwidth]{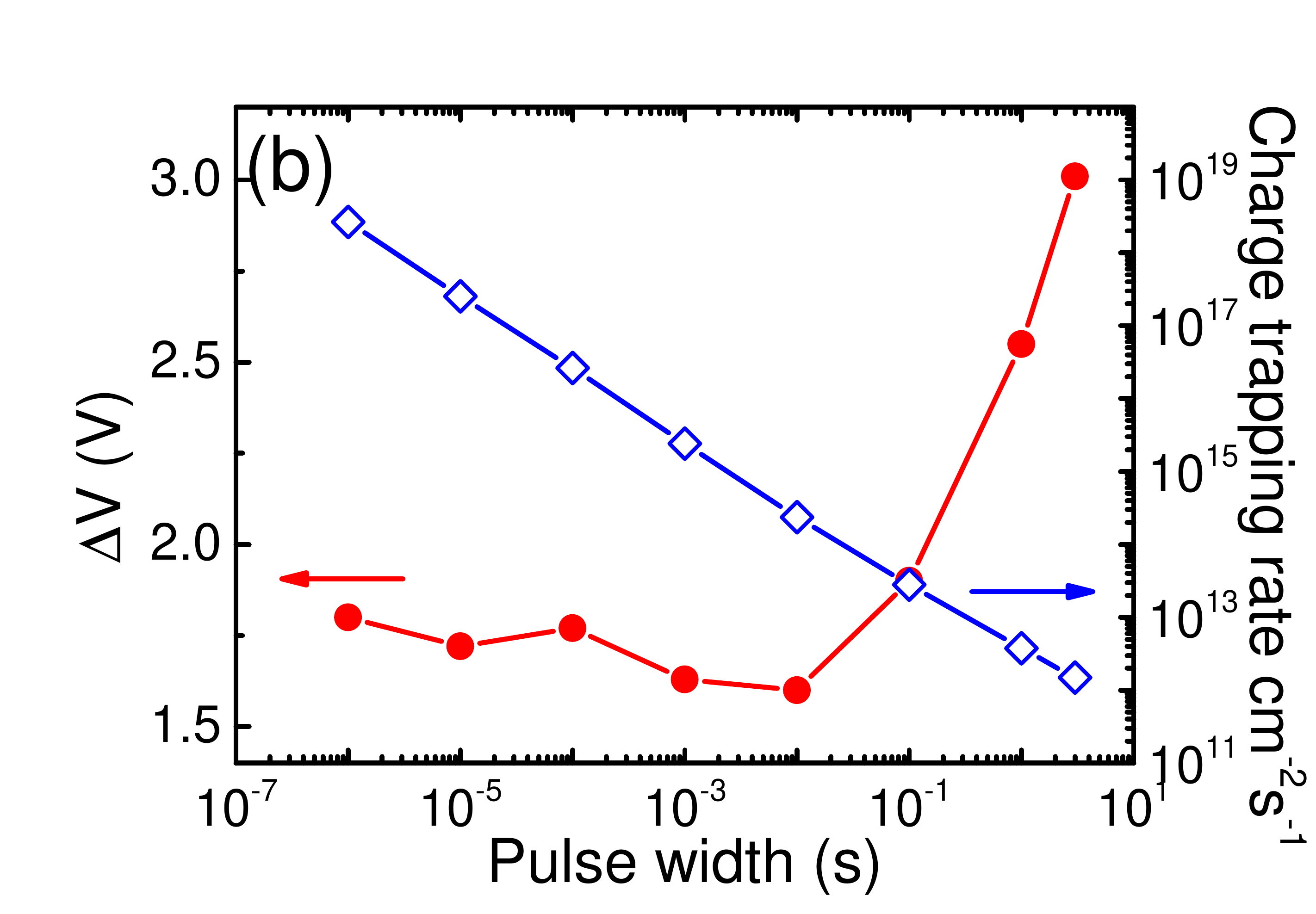}
	\caption{ (Color Online) {\bf  Dynamic Transition rate of the BP memory device} (a) Transfer curves of I$_{sd}$-V$_{TG}$ in a narrow range of +4 $\sim$ +12 V under different pulse duration (-16V, 1 $\mu$s to 3 s). (b) Extracted threshold voltage shift and calculated charge trap rate as a function of the pulse width, ranging between 1 $\mu$s and 3s.}\label{fig:charge-trap-rate}
\end{figure}

To study the dynamic transition rate of the black phosphorus memory device, a positive pulse (+16 V, duration of 3s) was applied to the top gate ($V_{BG}=0$) to set the device in the erase state, followed by a -16 V pulse with different duration time. The reading procedure was performed by sweeping I$_{DS}$-V$_{TG}$ in a very small range (+4 to +12 V) to minimize the effect of the measurements to the device's state.  After each reading operation, a positive pulse (+16 V, duration of 3s) was applied on the top gate to reset the device in the erase state. The threshold voltage shift $\Delta$V$_{TH}$ was acquired by applying in a linear fit to the linear regime of the reading I$_{DS}$-V$_{TG}$ curve. Figure \ref{fig:charge-trap-rate}(a) shows a clear shift of the threshold voltage when the width of the pulse is changed to 10 ms, which sets a reference for the following dynamic behaviour measurements. $\Delta$V$_{TH}$ shows nearly a saturation behaviour when the pulse width increases to 3 s (Figure \ref{fig:charge-trap-rate}(a)). The charge-trapping rate can be estimated from the expression:
\begin{equation}
	\frac{dN_{trap}}{dt} = \frac{C_{HF-AL}}{e} \times \frac{\Delta V_{TH}}{\Delta t}  \label{eqn:rate}
\end{equation}
where $\Delta$V$_{TH}$ is the threshold voltage shift and $\Delta$t is the pulse width \cite{2}. The calculated charge-trapping rate varies from 10$^{19}$ to 10$^{12}$ cm$^{-2}$t$^{-1}$ when the pulse width changes from 1 $\mu$s to 3 s. The reason for such a high charge-trap rate is because of the thin Al$_2$O$_3$ tunnel layer (only 5 nm), which makes electron/hole charges much easier to tunnel through.

\begin{figure}
	\centering
	\includegraphics[width=0.4\textwidth]{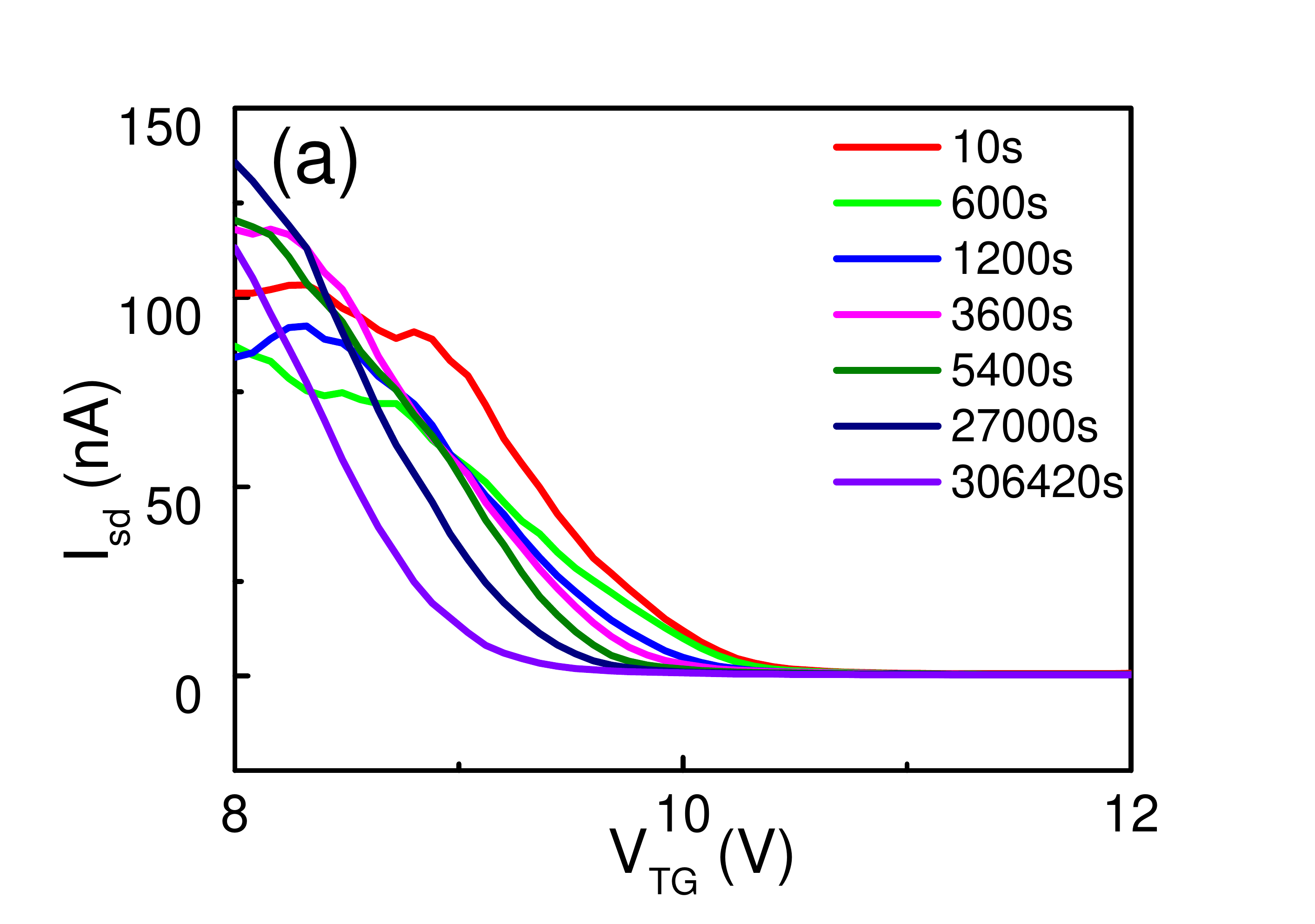}
	\includegraphics[width=0.4\textwidth]{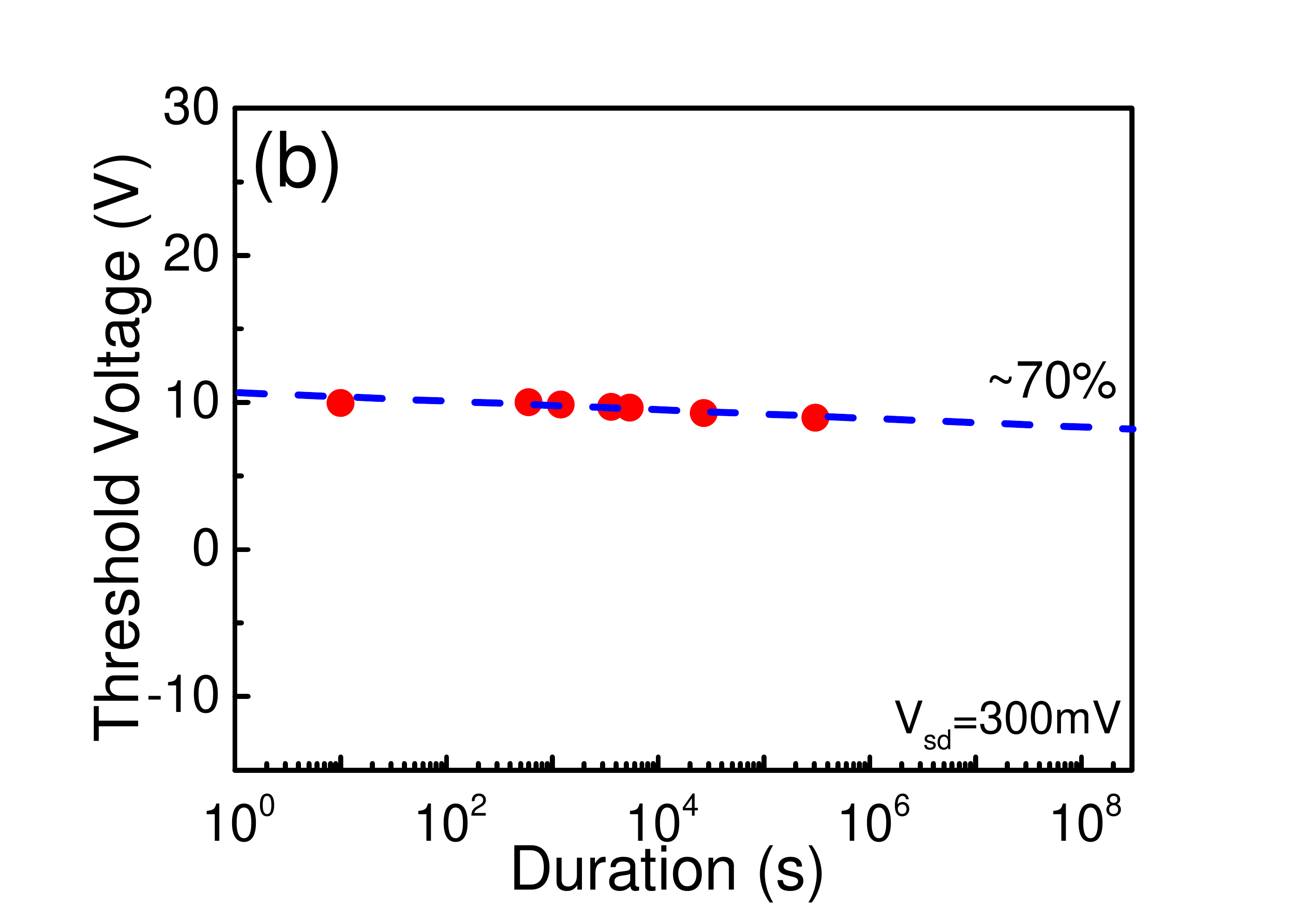}
	\caption{ {\bf  Retention characteristics of the BP memory device} (Color Online) { (a) The transfer curves (I$_{sd}$-V$_{TG}$) swept in different time intervals.(b) Retention time of the threshold voltage. The programing pulse is set to be -16V with 1s duration.  Threshold voltage was obtained by linearly fitting to the transfer curves. We estimate that only 30\% of the charges will be lost after 10 years. }}\label{fig:retention}
\end{figure}

The retention characteristics of the device are determined by the height of the tunneling barrier and the depth of potential well formed in the Al$_2$O$_3$/HfO$_2$/Al$_2$O$_3$ charge-trap stack \cite{39,50}. Figure \ref{fig:retention}(a) shows the threshold voltage at different time intervals after programming the device with a negative pulse (-16 V, duration of 1s). The transfer curve was also obtained in a small voltage range (+4  to +12 V). The extracted threshold voltage $\Delta$V$_{RTH}$ varies from 4.5 to 3.4 V after 10$^6$ s(Figure \ref{fig:retention}(b)), from which we estimate that only  30 $\%$ of the charges will be lost after 10 years. The enhancement of the retention  characteristics of our black phosphorus memory device is also related to the extraordinary trapping ability of the Al$_2$O$_3$/HfO$_2$/Al$_2$O$_3$ gate stack.

\begin{figure}
	\centering
	\includegraphics[width=0.4\textwidth]{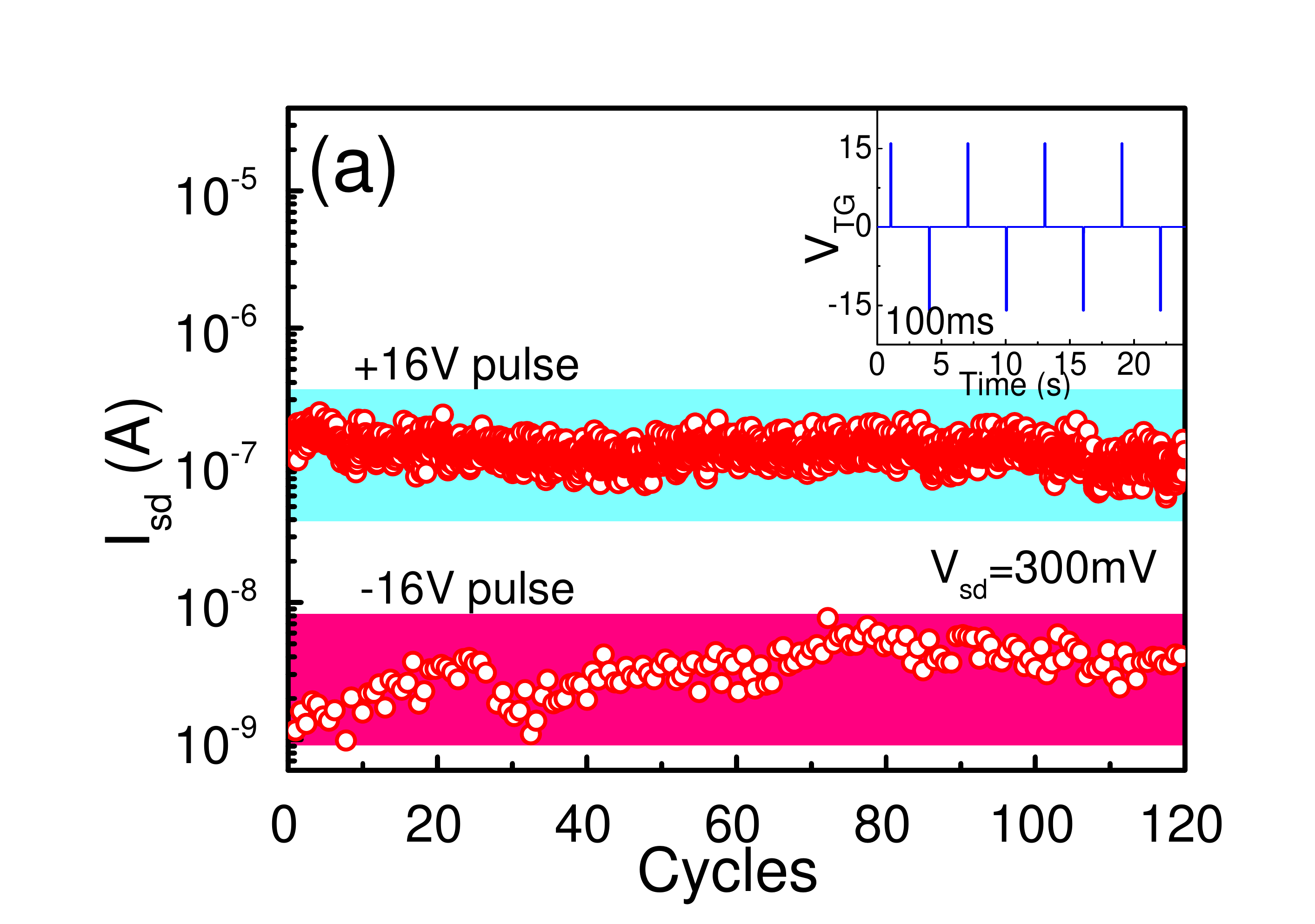}
	\includegraphics[width=0.4\textwidth]{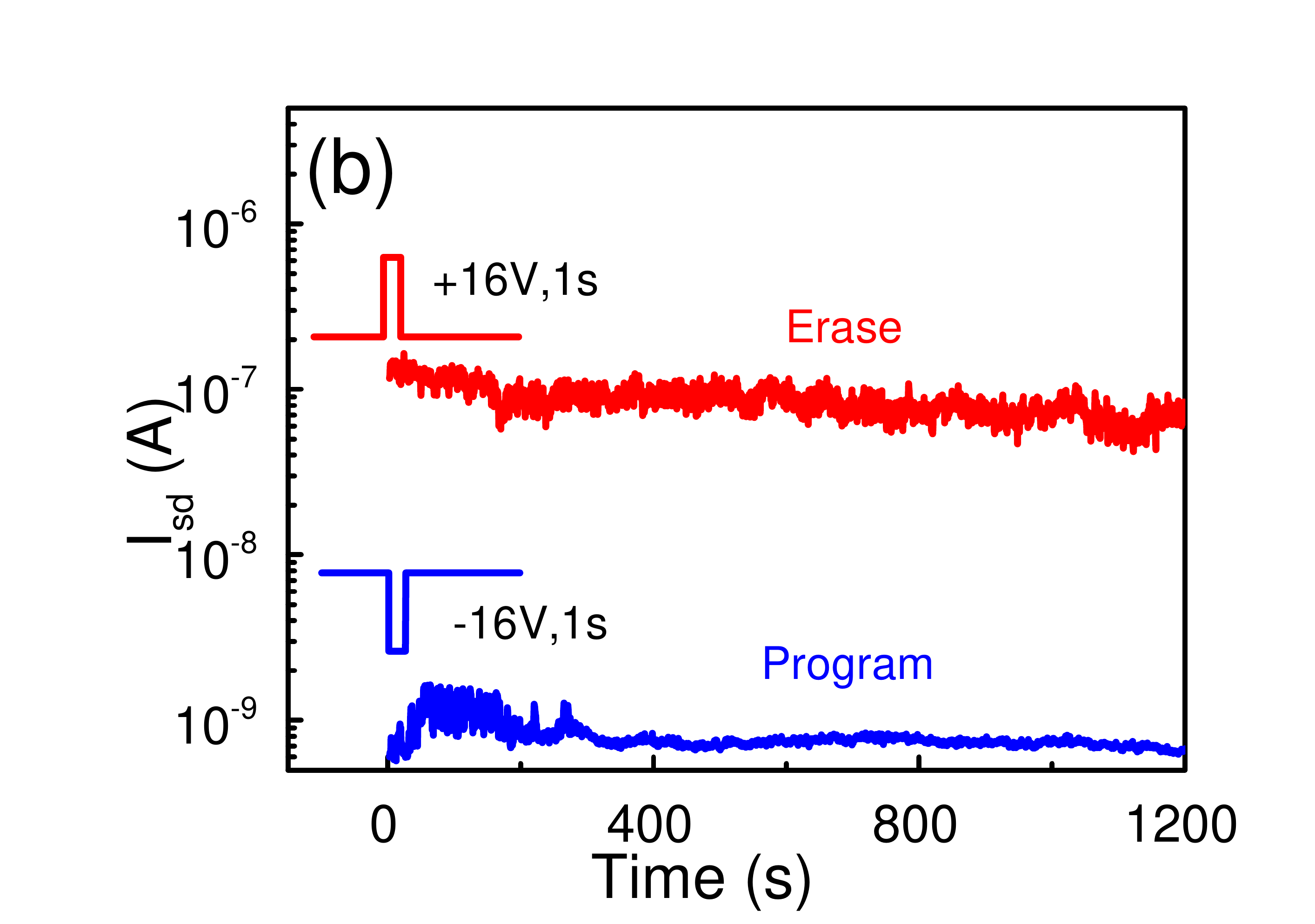}
	\caption{ (Color Online) {\bf  Endurance  of the BP memory device} (a) Endurance of the memory device for 500s, 120 cycles with the program/erase voltage being   +16, 100ms and -16 V,100ms. (b)The stability of program/erase state of the device after programming at +16V, 1s duration and erasing at -16V,1s (V$_{sd}$=300mV). The negative part of the threshold voltage shift correspondings to the hole trapping which can be suppressed by the positive back-gate voltages, vice versa for the electron trapping in the positive part.}\label{fig:endurance}
\end{figure}

\begin{table*}\scriptsize
	%
	\caption{\ Comparison of future memory device performance}
	\label{tbl:example}
	
	\begin{tabular*}{1\textwidth}{lllllll}
		\hline
		&Our work           & 2D Memory          & 2D memory          & 2D memory & Organic memory &   Organic memory\\
		&                   & MoS$_2$            & Graphene           & Graphene  &                &        \\
		
		\hline
		Channel material     & Black phosphorus &MoS$_2$              & Graphene           & Graphene       &    pentacene    & P3HT\\
		Charge trapping layer& HfO$_2$          &  Graphene           &MoS$_2$             &  GO            & Au nano-crystal  & Cu NPs\\
		&                  &  (few-layer)        &  (few-layer)       &                &monolayer&  \\    
		Barrier              & Al$_2$O$_3$      &  Al$_2$O$_3$/HfO$_2$&     h-BN           & Poly-vinylphenol&  Al$_2$O$_3$               &  PS or PVN\\
		Memory window        & 12V from $\pm$18V & 8Vfrom$\pm$15V       &  17Vfrom$\pm$40V  &  11.7Vfrom$\pm$20V& 2.25Vfrom$\pm$5V          & 42.6Vfrom$\pm$50V    \\
		Endurance            & >120             &$\sim$120            &  >100              & $\sim$ 200       & 1000 & >100  \\
		Retention (/10 years)& 30\%loss         & 70\%loss            & 70\%loss           & ----           & 40\%loss per day
		& 80\%loss\\
		Operating power      & V$_{sd}$=300mV   & V$_{sd}$=50mV       &V$_{sd}$=100mV       &V$_{sd}$=1.5V       &V$_{sd}$=3V& V$_{sd}$=20V        \\
		& Pulse<$\pm$16V   &  Pulse<$\pm$18V     & Pulse<$\pm$40V    &Pulse<$\pm$20V  &Pulse<$\pm$5V                 &Pulse<$\pm$50V\\
		Sample size          &  &                      &                    &               & &\\
		Lateral size         &   <5 $\mu$m&   <5 $\mu$m  & <5 $\mu$m   &              &  $\sim$1mm&inches\\
		Thickness &     $\sim$15nm    &$\sim$1nm  & <10nm                    &$\sim$1nm           &$\sim$30nm   &
		\\
		
		\hline
	\end{tabular*}\label{table1}
\end{table*}

The retention of trapped charge data was then studied. Two states with different currents can be defined as `trap' and `release' states, corresponding to program state and erase state, respectively. The retention performance curve could show that if the trapped charge can be maintained without loss of charge. In addition, as shown in Fig.\ref{fig:endurance}(b), high I$_{release}$/I$_{trap}$ ratio of 10$^2$ can be obtained, comparing with memory device with floating gate utilising 2D materials, e.g. GBM. Most of devices measured in this work exhibit similar retention characteristics regardless of thickness of charge-trap layer and tunnelling barrier.

To test the endurance of the memory device, a sequence of pulse ($\pm$ 16 V, duration of 100 ms) was applied to the top gate with V$_{BG}=0$ while I$_{DS}$ was measured (V$_{DS}=300$ mV). As presented in Figure \ref{fig:endurance}(a), the charge trapping can be preserved over 100 cycles. Most memory devices relying on charge trapping suffer from problems related to charge retention, such as loss of charge by back-tunnelling, injection of carriers of the opposite type or redistribution of charge in defects \cite{51}. The robustness and stability of the device shows a great perspective of applications in nonvolatile memory technology.

As 2D materials like graphene, MoS$_2$, even graphene oxide have received much attention in view of their application for nonvolatile memory, organic memories also have been studied as leading contending devices for future memory devices \cite{42,52,53,54,55}. Here, we compared the performances of our devices based on black phosphorus to those of reported memory devices based on other 2D materials and organic materials (Table \ref{table1}). Compared with graphene and MoS$_2$ charge trapping memories, the memory devices fabricated in our study display better performance with stable retention of 70\% retain after 10 years, while the memory window and endurance comparable to these devices. In addition, in terms of memory window, retention, miniaturization and low energy consumption, the device discussed in our work is obviously superior than those of organic memories. So it is reasonable to believe that nonvolatile memory devices based on black phosphorus hold great promise for future flexible and transparent memory devices. 

\section*{Conclusion}

In conclusion, we have demonstrated a  memory device based on few-layer black phosphorus and Al$_2$O$_3$ /HfO$_2$ /Al$_2$O$_3$  charge-trap stack. All the devices studied showed similar hysteresis characteristics, regardless of the channel length and BP film thickness. From the measurement of retention and endurance characteristics, it was confirmed that high-$k$ dielectric HfO$_2$ layer acts as an effective charge-trapping layer in this configuration. High on/off current ratio of 10$^2$, large memory window about 12 V at V$_{TG}$=16V, long data retention that only 30\% charge loss after 10 years, were attained. This study provides a promising route towrads the flexible and transparent memroy devices, utilising ultrathin two-dimensional materials.

\section*{Experimental section}
\subsection*{Device fabrication}

Thin layers of black phosphorus were mechanically exfoliated on a silicon wafer with 300 nm-thick SiO$_2$. The flakes were identified by a combination of optical and scanning electron microscopy (JEOL). We used multilayer BP with a thickness of $\sim$ 15 nm and widths of around $\sim$ 5  $\mu$m. The SiO$_2$/Si$^{++}$ is used as a back gate to control the carrier concentration in the BP. Source and drain electrodes of Ti (3 nm)/Au (85 nm) were fabricated  on BP samples by standard electron-beam lithography (EBL), followed by thermal evaporation and metal lift-off techniques. Electrodes with width 0.3-0.8 $\mu$m and a channel length of 0.5-2 $\mu$m were used. The wafers was covered with polymethyl methacrylate (PMMA) e-beam resist  immediately after   exfoliation to avoid possible degradation upon longer exposure to air. After lift-off, stacked layers of Al$_2$O$_3$/HfO$_2$/Al$_2$O$_3$ (5/8/35 nm) were deposited $via$ ALD. During the ALD process, trimethylaluminum and trtrakis (ethyl-methulamido) hafnium were reacted at 120 $^{\circ}$C with water for  Al$_2$O$_3$ and HfO$_2$, respectively. The top-gate (Ti/Au 3 nm/50 nm) electrodes were subsequently fabricated using another EBL and metal deposition process.

\subsection*{Characterization of BP material and device}

Atomic force microscopy (Bruker Multimode 8) was used to ensure the qualities and total film thickness of black phosphorus. Electrical properties of fabricated devices were measured with a semiconductor parameter analyzer (Agilent, B1500A) in vacuum and at room temperature.

\subsection*{Supplementary}

\begin{figure}[h]
	\centering
	\includegraphics[width=0.4\textwidth]{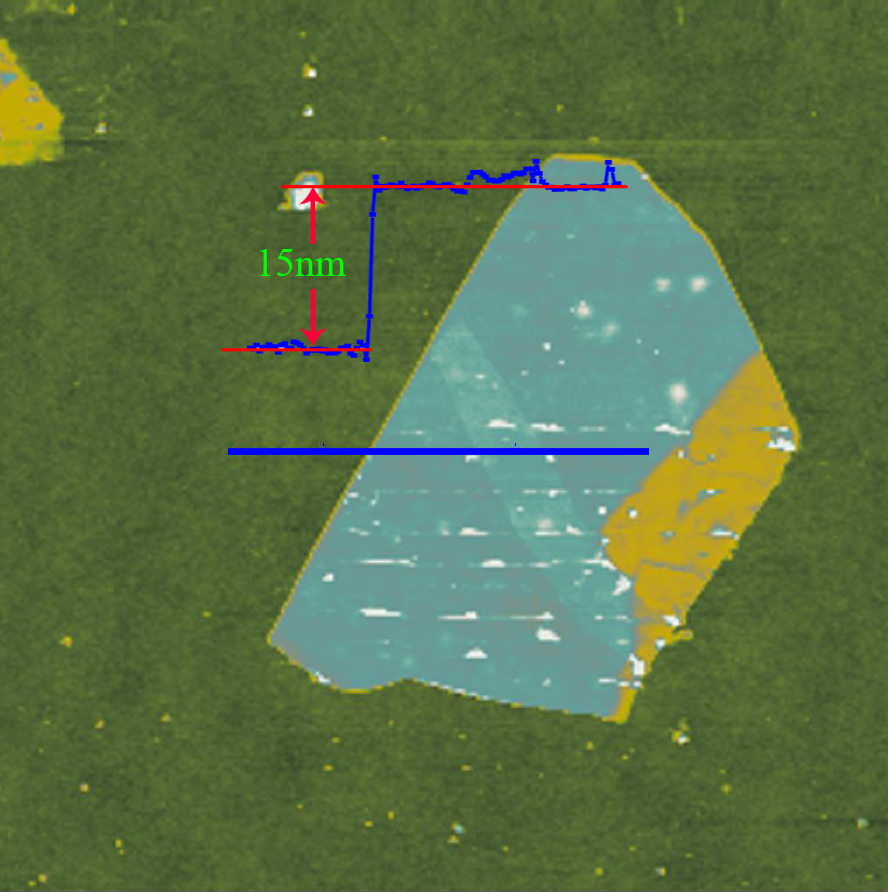}
	\caption{ (Color Online) {\bf  AFM measuements} }\label{fig:AFM}
\end{figure}

\begin{figure}[h]%
	\centering
	\includegraphics[width=0.4\textwidth]{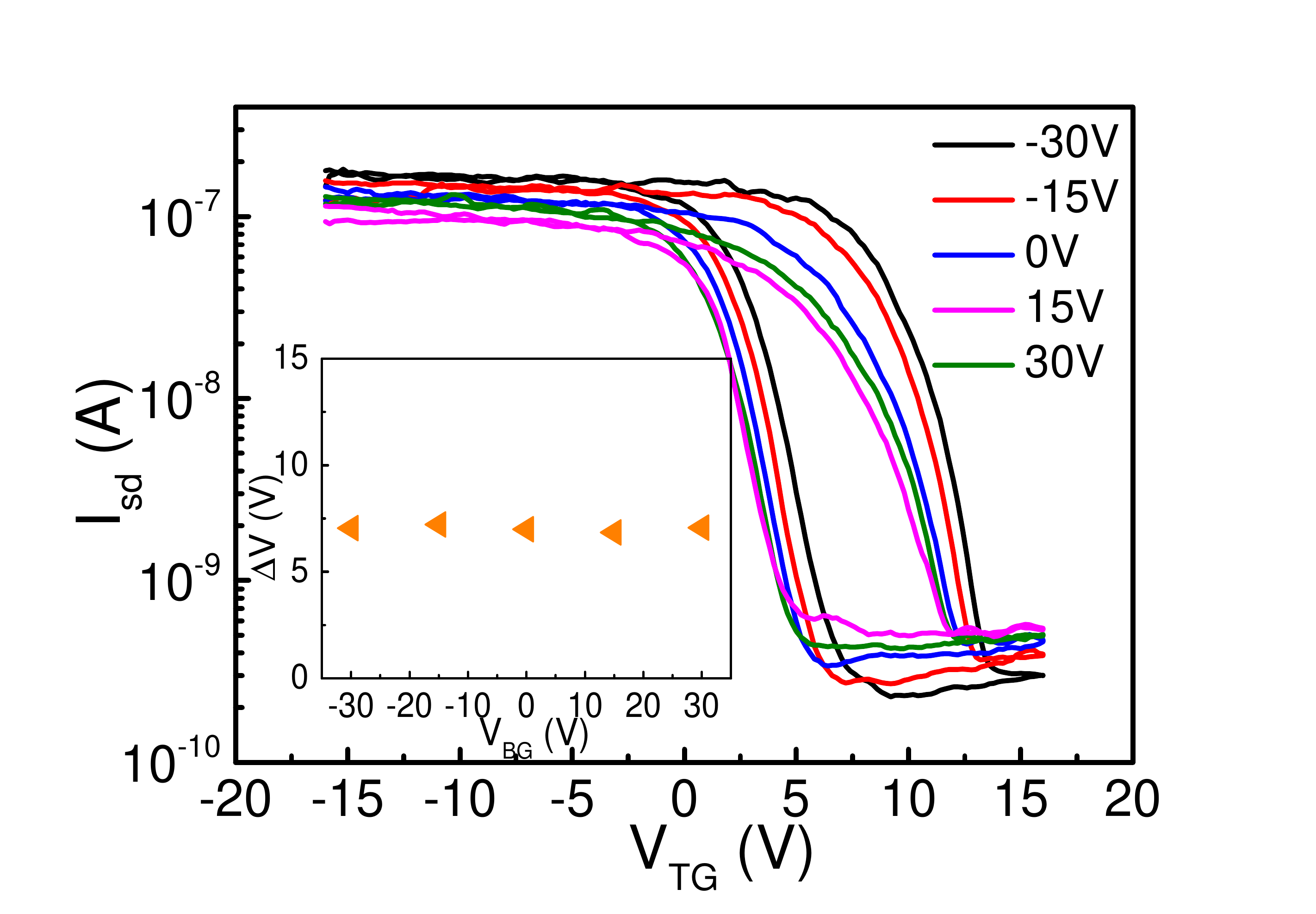} 
	\caption{ (Color Online) {\bf  Back-gate effect  of the BP memory device} }\label{fig:Vbg}
\end{figure}

\begin{figure}[h]
	\centering
	\includegraphics[width=0.4\textwidth]{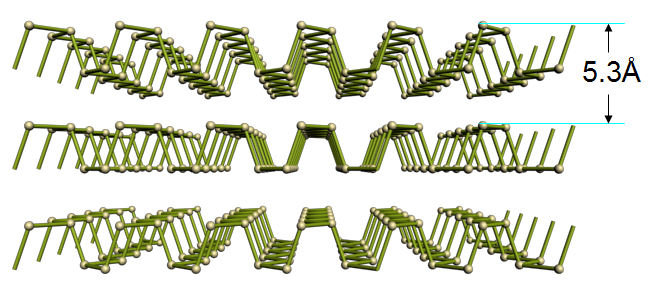}
	\caption{ (Color Online) {\bf  Device schematic} (a) Schematic of the memory device based on BP-AHA, (b) structure schematic of the black phosphorus. }\label{fig:structure}
\end{figure}

\begin{figure}[h]%
	\centering	
	\includegraphics[width=0.4\textwidth]{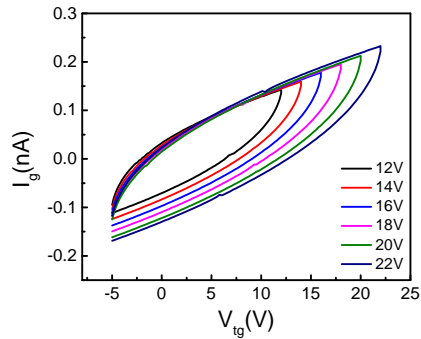} 
	\caption{ (Color Online) {\bf  Top-gate current leakage} Current leakage at different V$_{TG.MAX}$ }\label{fig:current-leakage}
\end{figure}

\section*{Conflict of Interest}

The authors declare no competing financial interest.

\section*{Acknowledgement}

This work was supported by '973 Program' Nos.2011CB922201, 2014CB643903, and NSFC Grant Nos.61225021, 11474272, and 11174272. The Project was Sponsored by the Scientific Research Foundation for the Returned Overseas Chinese Scholars, State Education Ministry.


\bibliographystyle{science}

\clearpage

\end{document}